\documentclass[12pt,preprint]{aastex}

\begin{document}

\title {The Chemical Composition of an Extrasolar Minor Planet}

\author{B. Zuckerman\footnote{Department of Physics and Astronomy and Center for Astrobiology, University of California, Los Angeles CA 90095-1562; ben, cmelis, hansen, jura @astro.ucla.edu}, D. Koester\footnote{Institut f\"{u}r Theoretische Physik und Astrophysik, University of Kiel, 24098 Kiel, Germany},\,\,\,\,C. Melis$^{1}$, Brad M. Hansen$^{1}$, \&  M. Jura$^{1}$}

\begin{abstract}
  We report the relative abundances of 17 elements in the atmosphere of the white dwarf star GD 362, material that, very probably, was contained previously in a large asteroid or asteroids with composition similar to the Earth/Moon system.  The asteroid may have once been part of a larger parent body not unlike one of the terrestrial planets of our solar system. 
\end{abstract}

\section{INTRODUCTION}

For white dwarfs with photospheric temperatures $<$25,000 K, elements heavier than hydrogen and helium sink out of the atmosphere in a time short compared to the time the star has spent as a white dwarf.  Consequently, discovery of metals, notably calcium, in the spectrum of numerous helium atmosphere (Sion et al. 1990a, 1990b; Koester et al. 2005b; Dufour et al. 2007 and references therein) and hydrogen atmosphere (Zuckerman et al. 2003; Koester et al. 2005b) white dwarfs stimulated a suite of explanatory models.  The models fall into three broad classes: fallback of dusty debris following a merger event (Hansen et al. 2006; Garcia-Berro et al. 2007), accretion of interstellar matter (see Koester \& Wilken 2006 for a recent discussion and additional references), and accretion of material from ancient planetary systems.   In the latter model, accreted metals were suggested to have their origin in comets (Alcock et al. 1986; Debes \& Sigurdsson 2002) or in asteroids (Jura 2003; Debes \& Sigurdsson 2002).  

In parallel with the optical/ultraviolet recognition of metal-bearing white dwarf photospheres, dusty orbiting disks were discovered at infrared wavelengths with ground-based telescopes (Zuckerman \& Becklin 1987; Kilic et al. 2005; Becklin et al. 2005) and with the Spitzer Space Telescope (Reach et al. 2005; Jura et al. 2007a, 2007b; von Hippel et al. 2007).  These dusty disks orbit some of the most metal-rich white dwarfs.  While the underlying cause(s) of the widespread presence of atmospheric metals is not yet fully understood, for many of the most heavily polluted white dwarf atmospheres, a promising model involves accretion of the heavy elements from dusty material whose origin was the tidal disruption of an asteroid (von Hippel et al. 2007, Jura et al. 2007a, and references therein).
  
Current insight into the composition of extrasolar material in the terrestrial planet zone in post-planet building eras has come from a tiny handful of stars that are orbited by large quantities of dust particles. Analysis of the 10-20 ${\mu}$m spectra of these stars reveals the nature of some silicate materials in the dust (Song et al. 2005; Beichman et al. 2005; Chen et al. 2006; Weinberger et al. 2007) but not the relative abundance of the elements.   In contrast, analysis of the spectra of the metal-polluted white dwarfs yields the relative abundances of individual elements, most commonly calcium, magnesium and iron.  If the relative rates at which various metals settle out of the white dwarf photospheres can be calculated, then the measured photospheric abundances provide a measure of the relative abundances of the elements in the accreted material.

In the present paper we focus on understanding the most dramatically polluted of all presently known white dwarfs, GD 362.   Gianninas et al. (2004) discovered, relative to hydrogen, an approximately solar abundance of calcium, magnesium and iron in the photosphere of GD 362.  Subsequently, the discovery of excess infrared emission from GD 362 demonstrated that it is orbited by circumstellar dust particles that are the likely source of the observed photospheric heavy elements (Becklin et al. 2005; Kilic et al. 2005).   Combining previous studies of this star with our new Keck/HIRES observations of 17 elements we are able to strongly confine the range of plausible models for the origin of the accreted metals; very probably they were contained in a large asteroid or asteroids with composition similar to the Earth/Moon system.  Thus, the outer atmosphere of GD 362 reveals the relative abundances of the elements in a body that was once part of an extrasolar planetary system.

\section{OBSERVATIONS AND ANALYSIS}
In June and September 2006 we obtained deep spectra of GD 362 with the HIRES echelle spectrometer (Vogt et al. 1994) on the Keck I telescope at Mauna Kea Observatory.  The spectra, of resolution ${\sim}$41,000, were analyzed over the range 3130-8670 {\AA}.   A much lower resolution LRIS (Oke et al. 1995) spectrum was obtained in September 2006 to aid in determination of overall atmospheric parameters.  We also analyzed a near-infrared spectrum over the range 10650-10810 {\AA} obtained in August 2006 with the NIRSPEC spectrometer (McLean et al. 2000) on the Keck II telescope.

We analyzed these various spectra with methods and input physics described in Finley et al. (1997) and Koester et al. (2001, 2005a, 2005b), and further references in these papers.  A surprise in the HIRES spectra is the presence of a line from He I at 5877.2 {\AA} (vacuum wavelength, Figure 1) and possibly at 6680.0 {\AA} where interpretation of the spectrum is difficult because of the presence of an Fe I line at essentially the same wavelength as the helium line.  While much weaker than the Balmer lines that initially implicated the photosphere of GD 362 as hydrogen-rich (Gianninas et al. 2004), the appearance of any He lines at all in a star as cool as ${\sim}$10,000 K heralds the presence of substantial quantities of helium.  This is illustrated in Table 1 where various combinations of atmospheric parameters are derived from the best fit of theoretical synthetic spectra to the LRIS profiles of the lowest four Balmer lines.  The quality of these fits is illustrated in Figure 2 where a grid with [He/H] = 10 was used.   In addition, our model fit for a pure hydrogen atmosphere is in excellent agreement with that of Gianninas et al. (2004).  In that case, we derive effective surface temperature (T$_{eff}$) and log gravity, $g$ (cm$^{2}$ s$^{-1}$) of 9850 ${\pm}$100 K and 9.09 ${\pm}$ 0.10, to be compared with 9740 ${\pm}$ 50 K and 9.12 ${\pm}$ 0.07 from Gianninas et al.

To best match the measured equivalent widths (EW) of the He 5877 and 6680 {\AA} lines (EW = 165 and upper limit of 30 m{\AA}, respectively), we interpolate in the fourth column of Table 1 to obtain T$_{eff}$, gravity, and ratio of helium to hydrogen abundance by number:
\begin{equation}
T_{eff}\; =\; 10540\, {\pm}\, 200 {\rm K},\;\;  \log g\; =\; 8.24\, {\pm}\, 0.04,\;\;  \log[\rm{He/H}]\; =\; 1.14\,{\pm}\, 0.10
\end{equation}
Thus, in actuality, the photosphere of GD 362 is helium, not hydrogen, dominated.  With the above parameters, and extensive line lists (Kurucz \& Bell 1995), we calculated synthetic spectra with approximately 2000 spectral lines of all elements in Table 2.  These synthetic spectra were then matched to the HIRES and NIRSPEC measured equivalent widths and elemental abundances (Table 2 and Figures 3 \& 4) were derived, as described in the paragraphs below.  This many elements have never been observed previously in the spectrum of any white dwarf.  The relative abundances of the elements is a balance between how rapidly each element is accreted onto the white dwarf from its surroundings and how rapidly each element sinks (diffuses) out of sight.  Calculations of the latter rates for some of the elements in Table 2 heavier than He indicate little difference in the timescales for sinking (Paquette et al. 1986; Koseter \& Wilken 2006).  Thus, in the following, we assume that the observed ratios in Table 2 are determined during the accretion process onto GD 362.   However, new diffusion calculations specifically tuned to the above parameters that characterize GD 362 and for the entire suite of elements listed in Table 2 are desirable.

We regard all Table 2 elements (other than C,N,O) to be securely detected in GD362 through multiple or unambiguous lines or, almost always, both.  Iron lines occur most commonly; we identify 232 in FeI and 47 in FeII.  Singly ionized titanium was second most prominent with 89 lines identified.   Elemental abundances were determined by comparison of measured equivalent widths (EW) with our calculated synthetic spectra.  As is well known, the increase of EW with abundance, the Òcurve of growthÓ, can be divided into three parts: the linear regime, where the EW depends only on the oscillator strength of the transition and linearly on the element abundance, the ÒflatÓ part, where the EW is almost constant, even with increasing abundance, and the Òsquare rootÓ part, where the EW increases with the square root of abundance, oscillator strength and damping constant.  Since the latter is often poorly known, for high precision abundances the linear regime is the preferred choice.  Unfortunately, even with our excellent HIRES spectra, we are not sufficiently sensitive to detect weak lines (a few m{\AA}) of the linear regime.  Thus, we analyzed lines in the square root regime ($>$100 m{\AA}), notwithstanding damping constant uncertainties.  Table 2 abundances are presented relative to H because H, not He, is the dominant opacity source and, also, for ease of comparison with solar abundance ratios. 

For each detected element observed lines were compared with those calculated from our synthetic spectra.  Varying the elemental abundances inputted into the synthetic spectra and comparing the predicted EW with those of multiple lines observed from a given element enabled estimation of abundance errors (see below).  The one exception is silicon for which we use only one line, at 3906 {\AA}, but it is clean and strong so the Si abundance error can be calculated reasonably. The synthetic spectral analysis made use of the Vienna Atomic Line Database (VALD).

The errors of the abundances were estimated as follows.  First we considered the scatter of abundances from fitting individual lines, and calculated an error of the mean from these.  In a second step, we added the uncertainty from fitting different ionization stages of one element; this was relevant only for Fe, Cr, Ca and Mg.  For those elements with very few and weak lines we have added an estimated contribution for the measurement error of the EW.  There is a slight remaining problem with the ionization balance, which seems to be systematic -- it is present for FeI/FeII, NiI/NiII, CrI/CrII, and MgI/MgII -- always in the sense that the ionic lines are predicted to be too strong.  For Mg the discrepancy is 0.2 dex, for the three others $<$0.1 dex.  These discrepancies disappear for a model with T$_{eff}$ = 10,240 K. 

No lines from C, N or O were detected.  The upper limits to abundances given in Table 2 were obtained from the HIRES spectrum from upper limits to the EW for N at ${\lambda}$ 8860 {\AA} and for O at ${\lambda}$ 7772/7774/7775 {\AA}.  For C, the abundance upper limit was derived from the NIRSPEC spectrum from a transition at ${\lambda}$ 10691 {\AA}. 

The measured (heliocentric) radial velocity of GD 362, including its gravitational red shift, is 49.3 ${\pm}$ 1.0 km s$^{-1}$.  With log g = 8.24 and white dwarf mass 0.73 M$_{\odot}$ (Section 3), the gravitational red shift is 43.3 km s$^{-1}$.

\section{DISCUSSION}

As may be seen, the abundance ratio to hydrogen of most of the elements in Table 2 is similar to the solar value.  What might this signify?  Because hydrogen never gravitationally settles out of a white dwarf's photosphere, the unknown portion of atmospheric hydrogen in GD 362 that accreted from the interstellar medium (Koester \& Wilken 2006) must have accumulated over the entire 10$^{9}$ year cooling age of the white dwarf.  In contrast, the atmospheric dwell time of Table 2 elements (other than helium) is about 100,000 years (Paquette et al. 1986).   Thus, if H and the heavy elements were accreted together from the interstellar medium over a very long period of time, then by now most of the heavy elements would have been removed from the photosphere.  Thus, unless most of the accretion of hydrogen from the interstellar medium has occurred during the past 100,000 years (even though the star is not now near any interstellar cloud), the observational result that the abundance ratios of most Table 2 elements to hydrogen are approximately solar must be coincidental. 

A combination of interstellar accretion to explain the hydrogen and circumstellar accretion to account for heavy elements appears to be the best model to explain the enormous range of calcium to hydrogen abundance ratios found in helium-rich white dwarfs (Sion et al. 1990a).  In addition, for the occasional DA that begins life as a white dwarf with a thin H atmosphere ($<$ 10$^{-13}$ M$_{\odot}$), transformation into a helium-rich atmosphere should be possible through convective mixing with an underlying He zone.  Such a possibility might apply to GD 362.

We now consider models for the origin of the heavy elements in the photosphere of GD 362 and conclude that, most plausibly, this origin was from accretion of an asteroid-like object or objects.  The asteroid(s) may well have initially derived from a parent body of planet-like mass.  If so, then we are measuring the composition of a portion of a present or former substantial planet (or moon) of GD 362. 

One possibility is that the material that accreted onto GD 362 came from interstellar space and had nothing to do with a former or current planetary system orbiting GD 362.  However, an interstellar origin is exceedingly unlikely for various reasons.  Jura et al. (2007b) argue that, because of a faint 24 ${\mu}$m flux, the overall mid-infrared spectrum of GD 362 indicates an outer edge to its dusty circumstellar disk; this is thus incompatible with the presence of cold dust that has drifted in toward the star from great distances.  Furthermore, the shape of GD 362's 10 ${\mu}$m silicate emission feature differs substantially from that of the silicate feature carried by interstellar dust particles (Jura et al. 2007b).  (These same two arguments pertain to the metal-polluted white dwarf G29-38 [Reach et al. 2005; Jura et al. 2007b].)   Figure 4 supplies an additional reason why the elements seen in GD 362 did not come from the interstellar medium.  With only a few exceptions (e.g., lithium) the composition of the solar photosphere is similar to that of the interstellar medium.  The solar (interstellar) ratio of calcium to sodium is of order unity and, as may be seen, this differs from the ratio in GD 362 by a factor of ${\sim}$30.  Also, the relative amount of carbon in the photosphere of GD362 is well below the interstellar value.  This carbon deficit relative to its interstellar abundance is similar to that deduced in three other externally polluted white dwarfs (Jura 2006).   Similar carbon deficits are found in some solar system bodies and are thought to represent the outcome of fractionation processes in the protosolar nebula.  All in all, elemental abundance ratios in GD 362 and in these other white dwarfs argue strongly against interstellar pollution models.

Another possibility that has been proposed (Garcia-Berro et al. 2007) attributes GD 362 and its circumstellar disk to a merger 2.2 Gyr ago of two white dwarfs. This scenario is not supported by observations of other candidate merged white dwarfs, which do not possess dust disks (Hansen et al. 2006), and because the abundance ratios in the atmosphere of GD 362 are very different from the nucleosynthetic predictions (Garcia-Berro et al. 2007).  In three different models, the disk-averaged value of n(C)/n(Ca) ranges between 100 and 750; we report an upper limit to this ratio of 5. The predicted values (Garcia-Berro et al. 2007) of n(Fe)/n(Mg) lie between 83 and 300; we measure a value of 2.

If, as argued above, stellar or interstellar events do not account for the heavy elements in the photosphere of GD 362, then to explain their origin we must turn instead to the possible roles played by various constituents of an ancient planetary system orbiting GD 362.  To evaluate these roles, it is important to understand the properties of the surface convection zone of GD 362.  White dwarfs are supported by electron degeneracy pressure.   However, in the very outer layers of a white dwarf, the material is non-degenerate, and, for cool white dwarfs, the dominant energy transport mechanism is convection. For mixed atmospheres with H/He ratios and effective temperatures like that of GD 362, the size of the convection zone lies somewhere in between that found for a pure hydrogen or a pure helium atmosphere at similar temperatures.  We calculate the size of the GD 362 convection zone using the evolutionary models of Hansen (1999), incorporating a boundary condition corresponding to a surface composition of hydrogen mass fraction X = 0.02, a total white dwarf mass of 0.73 M$_{\odot}$, and T$_{eff}$ = 10540 K.  We include in our mass estimates an order of magnitude uncertainty, to account for possible mixing with additional helium by convective overshooting (Freytag et al. 1996).   We estimate that the total mass of the convective zone ($>$98\% of which is helium) lies somewhere in the range 0.03 to one Earth mass.  Based on Table 2, nearly 0.01\% of the mass of the convective zone is contained in elements heavier than helium, specifically at least 10$^{22}$ g.  

This convective zone mass is of relevance to a model in which detected metals (Table 2) came from accretion of a comet or comets onto GD 362, either via direct hits, or via near misses that tear the comets apart with the resulting debris subsequently being accreted onto the star (Alcock et al. 1986; Debes \& Sigurdsson 2002).  All such comet accretion models are very unlikely.  Direct hits onto GD 362 of comets originating in an equivalent of the Sun's Oort comet cloud would occur about once every 10,000 years (Alcock et al. 1986) and the elements they deposit would remain in the convective atmosphere of GD 362 for about 100,000 years (Paquette et al. 1986) before they sink to the interior.  Thus, in this model, the residue of cometary material in the outer convective zone would amount to about 10 typical comets, or about 10$^{18}$ g (Zuckerman \& Becklin 1987; Festou et al. 1993).    Comparison of this mass with the above derived mass of metals in the convection zone ($>$10$^{22}$ g) indicates that direct comet hits fail by many orders of magnitude to explain the observations.

A model of tidal disruption of comets as they swing by the white dwarf also encounters multiple problems.  As mentioned above, the mid-infrared spectrum of GD 362 is incompatible with the presence of matter drifting in toward the star from great distances, as would be the case for comets originating in an equivalent of either the Kuiper Belt or Oort cloud; the former is located out beyond the orbit of Neptune, while the latter orbits the Sun at even much greater distances.  In addition, as may be seen in Figure 4, in the one comet in which a detailed comparison is possible, 81P/Wild 2 (Flynn et al. 2006), the relative abundances of certain elements, especially sodium and calcium, disagree substantially with the abundances in GD 362.

The mass in metals in the convective zone can be supplied by a large, icy, Kuiper Belt-like object that manages to survive the red giant phase of the evolution of GD 362.  However, like comets, initially such objects are expected to be composed, by mass, of ${\sim}$50\% of ices and perhaps to contain substantial amounts of carbon as well.  If so, then the oxygen abundance in GD 362, by number, would be about 3 times the total abundance of the refractory elements in Table 2.  But our measured upper limit to this ratio is only about unity (Table 2).  Thus, if the pollution of GD 362 is due to a large Kuiper Belt object, then most of its ice and probably carbon too must have been lost due to heating when GD 362 was a luminous red giant star. 

An alternative possibility is that elements seen in GD 362 initially belonged to an asteroid that wandered inside of the tidal radius of GD 362, was torn apart, and the resulting pieces eventually accreted onto the star (Debes \& Sigurdsson 2002; Jura 2003). Such a model is compatible with existing infrared measurements of GD362 (Jura et al. 2007b) and the metal mass in the convective zone can be supplied by one of the larger asteroids that orbit the Sun between Mars and Jupiter.

A question to consider is whether the polluting object originated in a debris belt -- akin, for example, to the Sun's asteroid belt or Kuiper Belt -- and was, thus, never a part of a planet-mass object or, alternatively, whether a planet or very large moon was the parent body for elements seen in GD 362.  Figure 4 illustrates the substantial difference in the relative abundances of sodium and calcium in CI chondrites (Lodders 2003) and in GD 362.  A similar discrepancy with GD362 abundances is apparent for other classes of meteorites and micrometeorites most all of which have [Na/Ca] ratios not much different than unity (e.g., Schramm et al. 1989).  The parent objects for these meteorites were generally not larger than asteroid-size.  In contrast, [Na/Ca] abundances in the Earth-Moon system show better agreement with those in GD 362 (Figure 3).  Thus, the [Na/Ca] ratio in GD 362 may be related to nebular condensation temperature or the evolutionary history of the parent body of origin, or both.  Given the Table 2 abundance errors, no general correlation of abundances and protoplanetary nebular condensation temperature (Appendix G in Wasson 1985) is apparent.  However, relative to nebular abundances, Na, with the lowest condensation temperature of Table 2 elements heavier than oxygen, appears to be significantly underabundant and calcium with an above average condensation temperature appears significantly overabundant.

The gravity given above for GD362 corresponds to a current mass of 0.73 ${\pm}$ 0.02 M$_{\odot}$, which, in turn, corresponds to a main sequence progenitor mass about three times that of the Sun.  The ratio of main sequence to giant star luminosity (when at maximum) is substantially greater for a 3 M$_{\odot}$ star than for the Sun.  As a consequence, inner rocky asteroids and planets that form at a given nebular temperature are more likely to survive the giant phase of stellar evolution around the 3 M$_{\odot}$ star.  Material in the innermost regions would likely be relatively deficient in volatiles like sodium.  In a planetary system structurally similar to our own, but around a three solar mass star, a Jupiter analog would orbit far from much of the inner region where material could survive through the red giant phase and where one might, therefore, expect only planets and not an asteroid belt.  However, if instead, a planetary system contained a warm Jupiter (as have been discovered elsewhere around some stars less massive than three solar masses), even in the inner regions rocky planet formation might be inhibited and a debris belt might exist.

In an ancient planetary system around GD 362, destabilized by mass loss from the red giant progenitor of GD 362 (Debes \& Sigurdsson 2002), a collision involving planets and/or large moons could have created a mix of interplanetary materials ranging in size from dust particles to asteroids.  The collision itself may have produced a large fragment with, by chance, an orbit that penetrated within the tidal destruction radius of GD 362.  Alternatively, subsequent to the initial collision, the gravitational field of the remnant of the planet might perturb an asteroid-size fragment toward GD 362.  A similar massive collision of two planet-size objects may have produced the huge quantities of dust seen in orbit around the main sequence star BD+20 307 (Song et al. 2005) and the shapes of the very strong silicate emission features at BD+20 307 and GD 362 are quite similar (Jura et al. 2007b).

\section{CONCLUSIONS}

The photosphere of GD 362 reveals a suite of elements many of which have never before been detected in the atmosphere of a white dwarf, including, for example, strontium and scandium (Figures 5 \& 6) whose abundances are nine orders of magnitude less than that of hydrogen.   We conclude that tidal disruption of an asteroid or asteroids with composition similar to that of the Earth/Moon system is the most plausible explanation of the observations.  If so, then the photosphere of GD 362 has provided astronomers with the first comprehensive measurement of the bulk elemental composition of a body that was once a part of an extrasolar planetary system.  Additional insight should come when an ultraviolet spectrometer above Earth's atmosphere once again becomes available to astronomers thus enabling measurement of elemental abundances in GD 362 beyond those listed in Table 2.

We are very grateful to A. Rau, E. Ofek, and S. Kulkarni for obtaining a LRIS spectrum of GD 362 and to J. Wasson for helpful suggestions.  We thank referee Pierre Bergeron for constructive comments.  This research was supported in part by NASA grants to UCLA.

\newpage
\begin{center}
Table 1: Model parameters for GD 362 as a function of [He]/[H] by number
\\
\begin{tabular}{llllllll}
\hline
\hline
He/H & T$_{eff}$ & $\log$ $g$ & EW (He I 5877)\\
    &    (K) & (cm$^{2}$ s$^{-1}$) & ({\AA}) \\
 \hline
 0.1	&	10000	&	9.00	&	0.000\\
		1.0	&	10010	&	8.66	&	0.070\\
		10		&10340	&	8.28	&	0.120\\
		100	&	11860	&	7.98	&	1.270 \\
  \hline
 \end{tabular}
 \end{center}
 
\newpage
\begin{center}
Table 2: Elemental Abundances by number relative to hydrogen in GD 362
\\
\begin{tabular}{llllrlrrrr}
\hline
Element & log [element/H] & Solar (Lodders 2003)  \\
\hline
           He	&	1.14 ${\pm}$ 0.10	&	-1.10 ${\pm}$ 0.01\\
			C	& 	$<$-4.50		&	-3.61 ${\pm}$ 0.04\\
			N	&	$<$-3.00	&		-4.17 ${\pm}$ 0.11 \\
			O	&	$<$-4.00	&		-3.31 ${\pm}$ 0.05 \\
			Na	&	-6.65 ${\pm}$ 0.20	&	-5.70 ${\pm}$ 0.03 \\
			Mg	&	-4.84 ${\pm}$ 0.25	&	-4.45 ${\pm}$ 0.02\\
			Al		&-5.26 ${\pm}$ 0.20		&-5.54 ${\pm}$ 0.02\\
			Si		&-4.70 ${\pm}$ 0.30		&-4.46 ${\pm}$ 0.02\\
			Ca	&	-5.10 ${\pm}$ 0.10	&	-5.66 ${\pm}$ 0.03\\
			Sc		&-9.05 ${\pm}$ 0.30	&	-8.93 ${\pm}$ 0.04\\
			Ti		&-6.81 ${\pm}$ 0.10	&	-7.08 ${\pm}$ 0.03\\
			V		&-7.60 ${\pm}$ 0.30	&	-8.00 ${\pm}$ 0.03 \\
			Cr		&-6.27 ${\pm}$ 0.10	&	-6.35 ${\pm}$ 0.05\\
			Mn	&	-6.33 ${\pm}$ 0.10	&	-6.50 ${\pm}$ 0.03\\
			Fe		&-4.51 ${\pm}$ 0.10	&	-4.53 ${\pm}$ 0.03\\
			Co	&	-7.36 ${\pm}$ 0.40& 		-7.09 ${\pm}$ 0.03\\
			Ni		&-5.93 ${\pm }$ 0.15	&	-5.78 ${\pm}$ 0.03\\
			Cu	&	-8.06 ${\pm}$ 0.40	&	-7.74 ${\pm}$ 0.06\\
			Sr		&-9.28 ${\pm}$ 0.30 & -9.09 ${\pm}$ 0.04\\

\hline
\end{tabular}
\end{center}
\newpage
\begin{figure}
\plotone{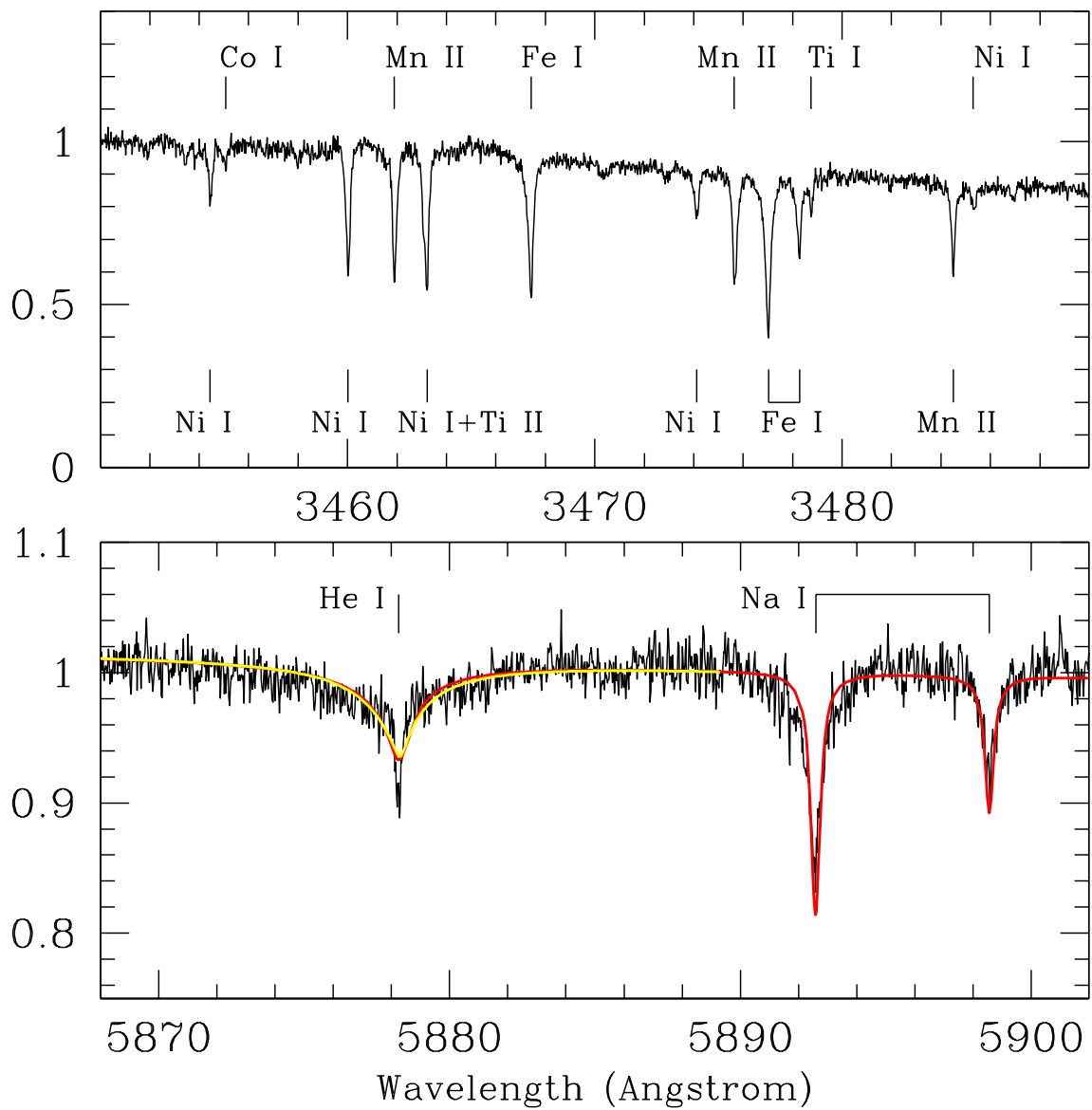}
\caption {Representative HIRES spectra.  Wavelengths are in vacuum and a heliocentric
rest frame.  The ordinate of the upper panel is in arbitrary units and is not flux calibrated.  The spectrum in the lower panel has been corrected for the instrumental response.  Our model atmosphere spectra (colored lines) do not match the depth of the narrow core of the helium line seen in the lower panel.  Perhaps non-LTE effects, similar to those seen in the cores of H${\alpha}$ lines in DA white dwarfs, are present.  The red and the yellow lines correspond to two different line broadening theories.}
\end{figure}
\begin{figure}
\plotone{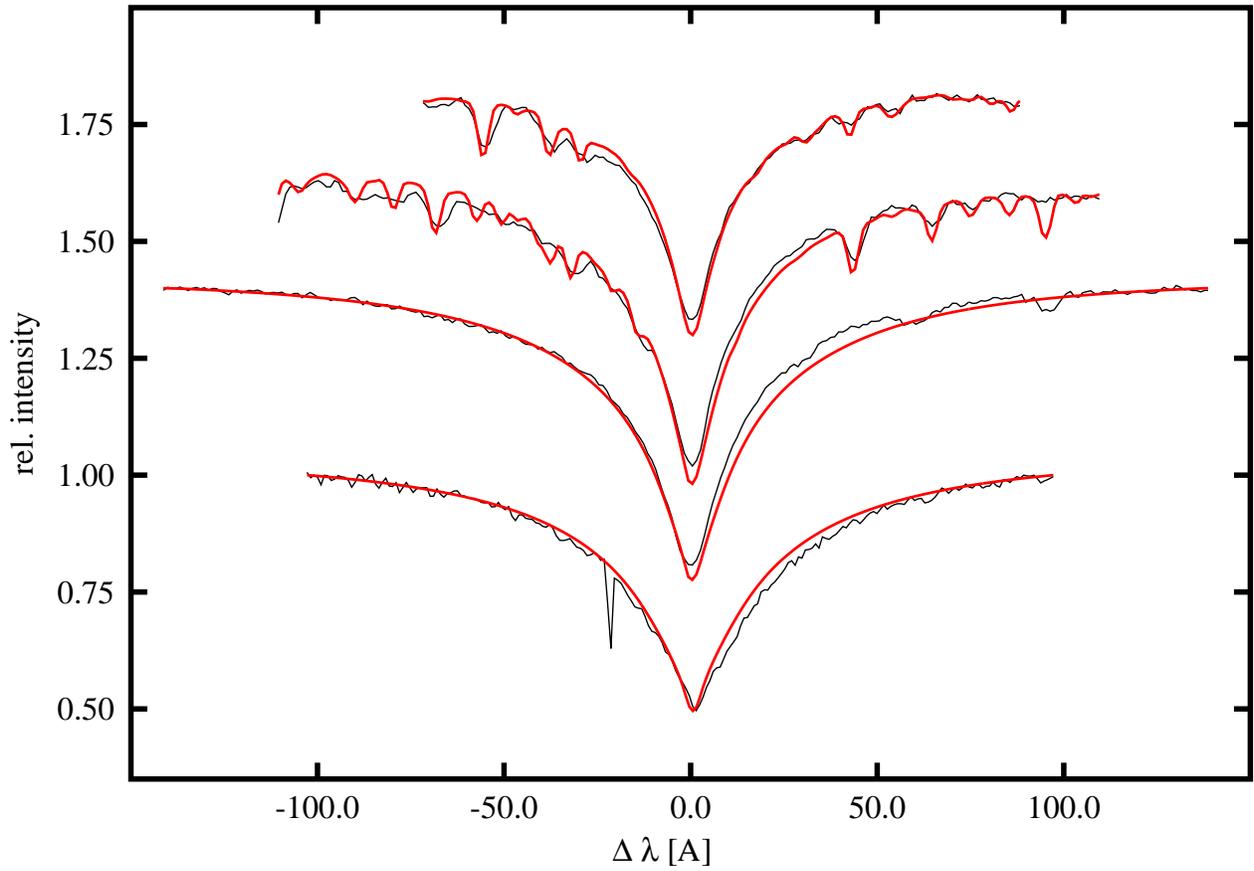}
\caption { Best fit models (in red) for the lowest four Balmer lines based on a grid with [He/H] = 10 by number and including metals with approximately final abundances.  The underlying spectrum in black is from LRIS.}
 \end{figure}
 \begin{figure}
 \plotone{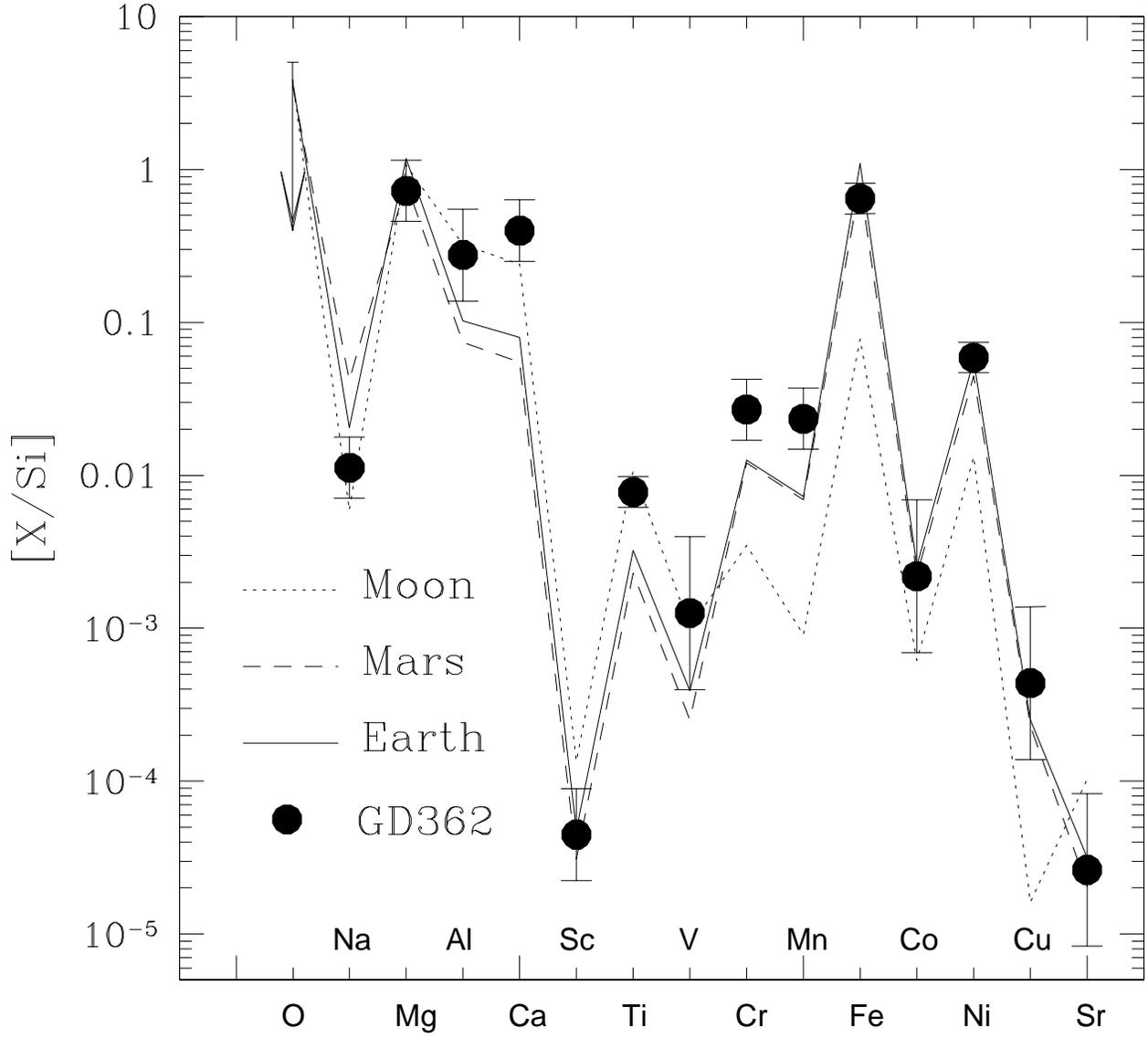}
 \caption{Elemental abundances by number relative to silicon.  GD 362 data from Table 2, Earth, Moon and Mars from Lodders \& Fegley (1998). }
 \end{figure}
 \begin{figure}
 \plotone{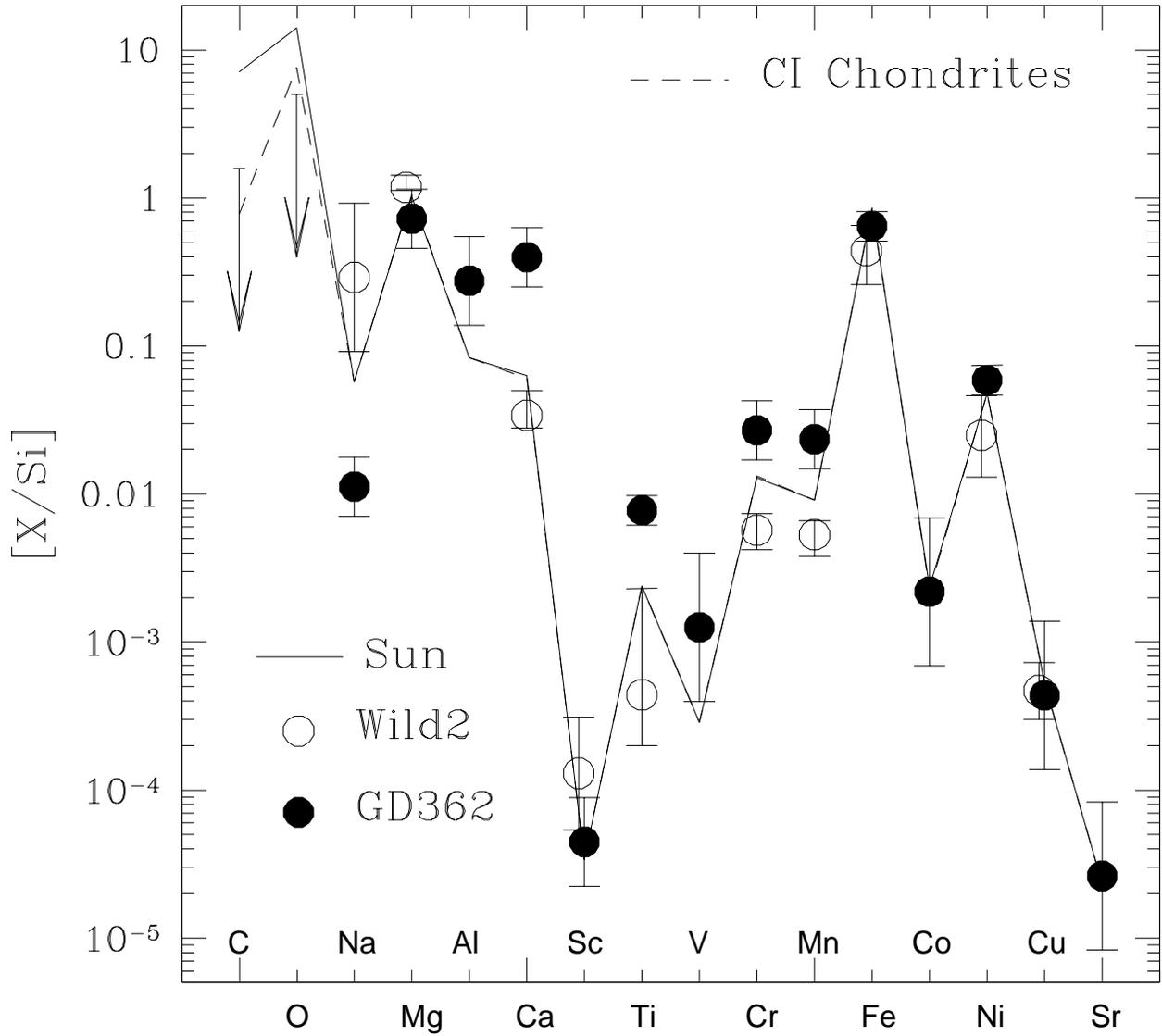}
\caption{Elemental abundances by number relative to silicon.  GD 362 data from Table 2,
with solar and CI chondrite abundances from Lodders (2003), and 81P/Wild 2 from Flynn et al (2006).}
 \end{figure}
 \begin{figure}
\plotone{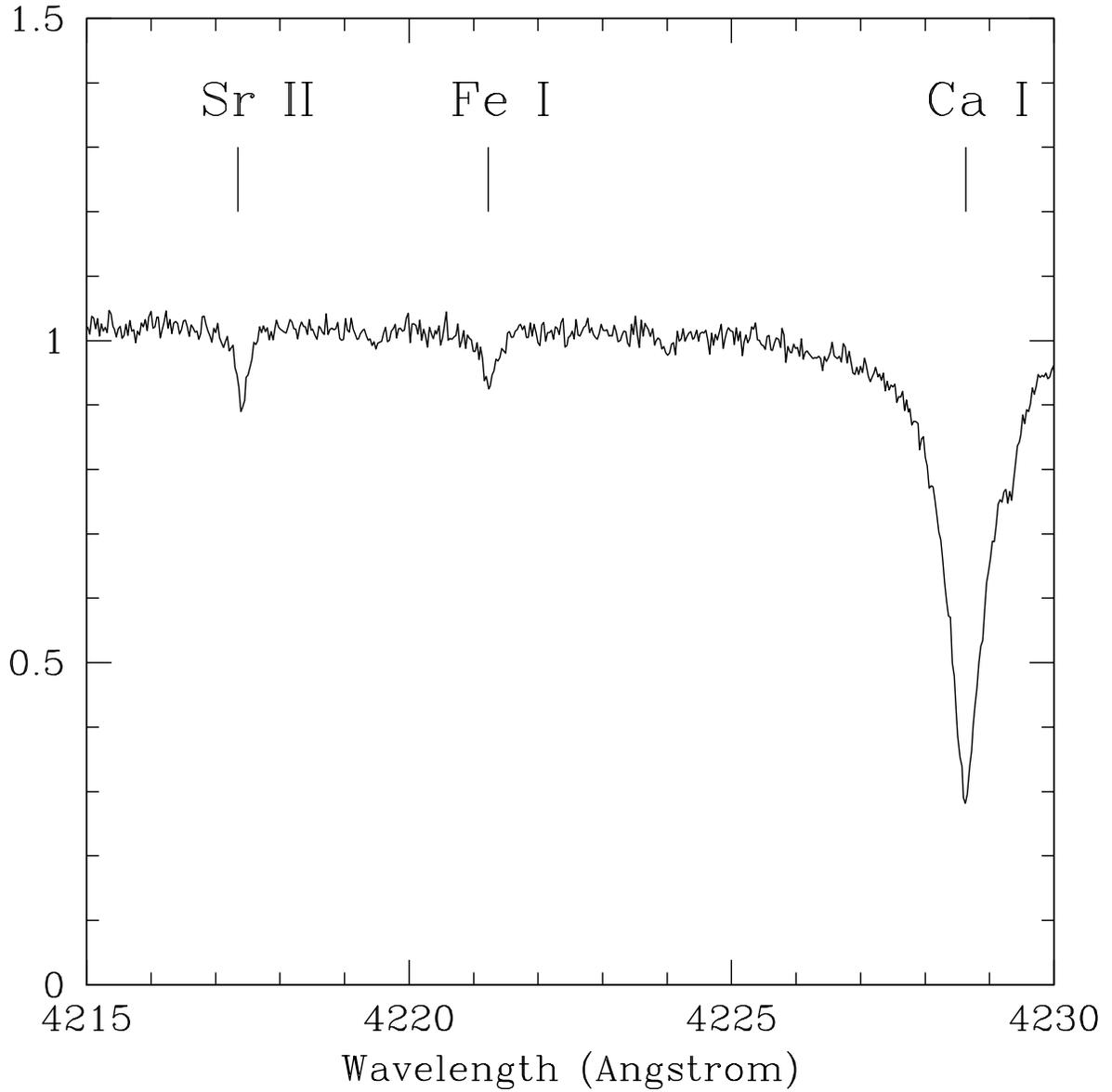}
 \caption{ Representative HIRES spectrum including a line from strontium whose abundance in the Sun and in GD 362 is nine orders of magnitude less than that of hydrogen. Wavelengths are in vacuum and a heliocentric rest frame.  The ordinate is in arbitrary units and is not flux calibrated. }
\end{figure}
 \begin{figure}
 \plotone{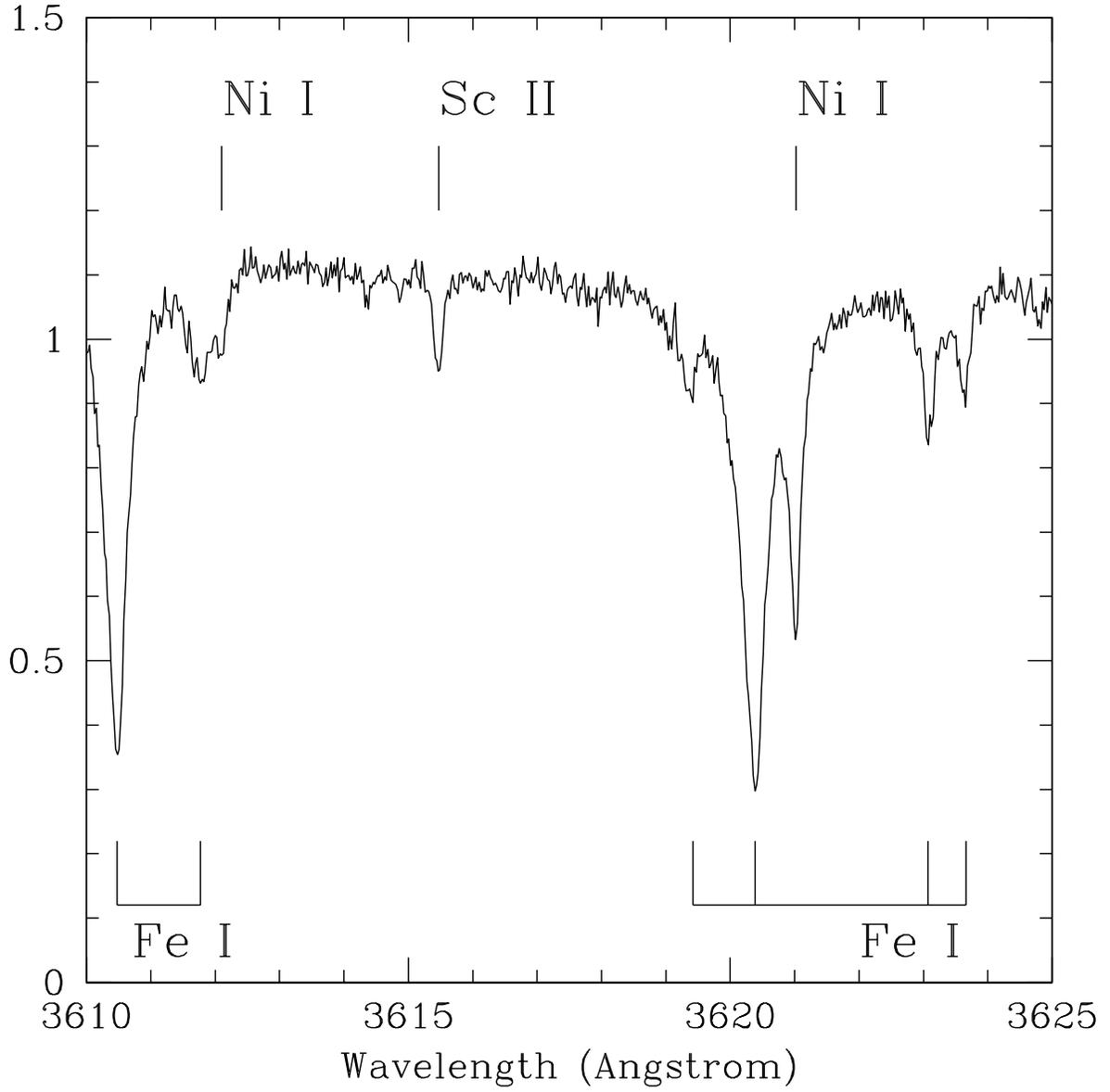}
 \caption{ Same as Figure 5, but for scandium. }
 \end{figure}
\end{document}